\documentclass{mn2e}

\usepackage{graphicx}
\usepackage{amssymb}
\usepackage{txfonts}

\newcommand{\sax}{{\it Beppo\-SAX}}
\newcommand{\msun}{{\rm M}_{\sun}}
\newcommand{\rsun}{{\rm R}_{\sun}}

\title[Stellar wind and X-ray spectra of Cygnus X-3]{Effects of the stellar wind on X-ray spectra of Cygnus X-3}

\author[A. Szostek and A. A. Zdziarski]
{
Anna Szostek\thanks{E-mail: asz@camk.edu.pl, aaz@camk.edu.pl}
and Andrzej A.~Zdziarski\footnotemark[1]\\
Centrum Astronomiczne im.\ M. Kopernika, Bartycka 18, 00-716 Warszawa, Poland\\
}

\date{Accepted 2008 February 04. Received 2008 January 25; in original form 2007 October 30}

\pagerange{\pageref{firstpage}--\pageref{lastpage}}
\pubyear{2008}

\begin{document}

\maketitle

\label{firstpage}

\begin{abstract}
We study X-ray spectra of Cyg X-3 from \sax, taking into account absorption and emission in the strong stellar wind of its companion. We find the intrinsic X-ray spectra are well modelled by disc blackbody emission, its
upscattering by hot electrons with a hybrid distribution, and by Compton
reflection. These spectra are strongly modified by absorption and reprocessing in the stellar wind, which we model using the photoionization code {\tt cloudy}. The form of the observed spectra implies the wind is composed of two phases. A hot tenuous plasma containing most of the wind mass is required to account for the observed features of very strongly ionized Fe. Small dense cool clumps filling $\la$0.01 of the volume are required to absorb the soft X-ray excess, which is emitted by the hot phase but not present in the data. The total mass-loss rate is found to be (0.6--$1.6)\times 10^{-5}\msun$ yr$^{-1}$. We also discuss the feasibility of the continuum model dominated by Compton reflection, which we find to best describe our data. The intrinsic luminosities of our models suggest that the compact object is a black hole.
\end{abstract}
\begin{keywords}
accretion, accretion discs -- binaries: general -- stars: individual: Cyg X-3
--  X-rays: binaries -- X-rays: individual: Cyg X-3 -- X-rays: stars. 
\end{keywords}

\section{Introduction}

Cyg X-3 is a high-mass X-ray binary (XRB) system with a short orbital period, $P=4.8$ h. It is located at a distance of $d\sim 9$ kpc in the Galactic plane (Dickey 1983, assuming the 8 kpc distance to the Galactic Centre; Predehl et al.\ 2000). Due to strong interstellar absorption, its optical counterpart is not observable, though infrared observations indicate it is a Wolf-Rayet (WR) star (van Kerkwijk et al.\ 1996). In spite of its discovery already in 1966 (Giacconi et al.\ 1967), it remains poorly understood. In particular, it remains uncertain whether its compact object is a black hole or a neutron star. Currently, there are two other known X-ray binaries containing WR stars, IC 10 X-1 (Prestwich et al.\ 2007) and NGC 300 X-1 (Carpano et al.\ 2007a, b). In both cases, there is dynamical evidence that the compact object is a black hole. 

Cyg X-3 is the brightest radio source among X-ray binaries (Mccollough et al.\
1999). It shows very strong radio outbursts and resolved jets (Marti, Paredes \& Peracaula 2000; Mioduszewski et al.\ 2001). In the X-rays, it exhibits a wide range of variability patterns. In particular, transitions between the hard and soft spectral state occur on the timescale of months to years. 

The understanding of the radiative processes underlying the X-ray spectra of Cyg X-3 remains rather rudimentary. Zdziarski \& Gierlinski (2004) have shown an overall similarity of the spectral states of Cyg X-3 to canonical states of black hole binaries. However, details, in particular the energy of the cutoff in the hard state, appear to differ significantly. Many different models have been fitted to spectra of Cyg X-3 by various authors, but each of them appeared to fit the data well, which precluded determination of the correct one. The models assumed different system components, geometry, radiative processes and the absorbing medium. They included different combinations of absorbers, Gaussian lines, blackbody, power-law with or without a cutoff, and bremsstrahlung, and different model combinations were proposed for the hard and soft states, see, e.g., White \& Holt (1982), Nakamura et al.\ (1993), Rajeev et al.\ (1994). Later, the power-law models were replaced by more physical models of Comptonization, see Vilhu et al.\ (2003), Hjalmarsdotter et al.\ (2004) and Hjalmarsdotter et al.\ (2008, hereafter H08). The interpretation of the intrinsic unabsorbed spectra, remains, however, not clear. In particular, H08 show that {\it INTEGRAL\/} data of Cyg X-3 in the hard state could be well modelled by three models with very different shapes of the intrinsic continuum,  yet the model most promising from a physical point of view yielded the worst statistical fit.

A key issue in the search for understanding the spectra of Cyg X-3 appears to be the nature of the complex intrinsic absorption, most likely caused by the wind from the companion star. In this paper, we study the effects of the WR stellar wind on the X-ray spectra. In Section \ref{data}, we present the data used for our study. Then, we discuss our theoretical models in Section \ref{s:model}. In Section \ref{results}, we describe the complex interactions between the X-rays and the wind, and present results of our fits to the data. Sections \ref{discussion} and \ref{conclusions} contain a discussion of our results and our conclusions, respectively. 

\section{The data}
\label{data}

The \sax\/ satellite carried on board four co-aligned instruments, the
Low-Energy Concentrator Spectrometer (LECS; Parmar et al.\ 1997), the Medium-Energy Concentrator Spectrometers (MECS; Boella et al.\ 1997), the High-Pressure Gas Scintillation Proportional Counter (HPGSPC; Manzo et al.\ 1997), and the Phoswich Detector System (PDS; Frontera et al.\ 1997). Together, they covered the nominal energy range of 0.1--200 keV,  

We choose here two out of the nine Cyg X-3 observations carried out by \sax, one in the soft state (hereafter SS), and one in the hard state (hereafter HS). The log of the observations is given in Table \ref{log}. The LECS, MECS and PDS event files were obtained from the on-line public archive of the \sax\/ Science Data Centre (hereafter SDC), whereas the HPGSPC files for the SS observation have been generated by the \sax\/ SDC for us. The MECS data are available from the units MECS2 and MECS3, which we add together. There are no HPGSPC data for the HS observation. Following the recommendations of the SDC, we have used the energy bands for the instruments as follows: the LECS, 0.12--4.0 keV, the MECS, 1.65--10.0 keV, the HPGSPC, 7.0--34 keV, and the PDS, 15--220 keV. 

\begin{figure}
\centerline{\includegraphics[width=80mm]{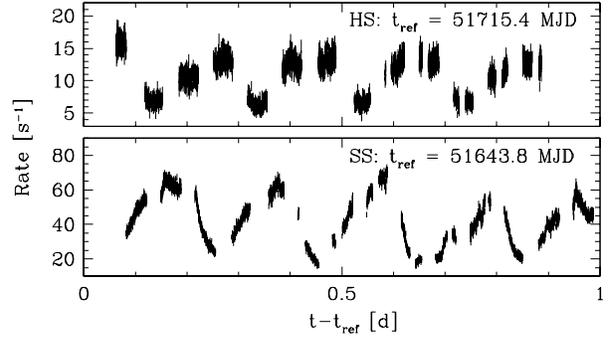}}
\caption{The background-subtracted MECS2 1.65--10.5 keV light curves. The orbital modulation at $P\simeq 1/5$ d is clearly seen.
} 
\label{lc}
\end{figure}

The LECS and MECS data have been selected from circular regions centred on the
source with an $8'$ radius. We have used blank field observations to obtain the
MECS background. Because Cyg X-3 is located at a low Galactic latitude of
$+0.8\degr$, and in a crowded region, we have used the semi-annuli method
(Parmar et al.\ 1999) to obtain the LECS background. The background for the two
remaining, high-energy and non-imagining, instruments come from off-source
observations. The MECS2 light curves are shown in Fig.\ \ref{lc}. We clearly see strong orbital modulation, characteristic to this source. 

\begin{table}
\centering
\caption{The log of the analysed \sax\/ observations.}

\begin{tabular}{ccccc}
\hline
ObsID      & Instrument & Start date        & Exposure   & Count rate \\
           &            & Start time [UT]   &  [ks]  & [s$^{-1}$] \\
\hline

\multicolumn{5}{c}{Soft State}\\
209530012  & LECS       & 2000 April 9  & 11.8 & $23.08 \pm 0.04$ \\
           & MECS       & 21:02:10.5   & 81.0 & $14.65 \pm 0.01$ \\
           & HPGSPC     &             & 36.0 & $12.10 \pm 0.08$ \\
           & PDS        &             & 18.0 & $2.26  \pm 0.04$ \\
\multicolumn{5}{c}{Hard State}\\
209530015  & LECS       & 2000 June 20 & 17.5 & $4.16 \pm 0.02$  \\
           & MECS       & 11:05:47.5   & 68.9 & $5.77 \pm 0.10$  \\
           & PDS        &             & 14.0 & $28.20 \pm 0.07$ \\
\hline
\end{tabular}
\label{log}
\end{table}

The spectra (which are averaged over about 5 orbital cycles) have been grouped to satisfy the minimum number of counts in each channel as required by the $\chi^2$ statistic. The LECS and MECS channels have been grouped up to the minimum of 40 counts per bin. The HPGSPC data are grouped by the instrument software according to the signal-to-noise ratio. The PDS spectra have been grouped into 18 bins with different minimum number of counts according to the SDC recommendations. 

We then fit simultaneously all of the available instruments. We have allowed free normalization of the LECS and HPGSPC data with respect to the MECS data, and the fitted values were about 0.9--1 for the LECS, and 0.97--1.05 for the HPGSPC, which is in good agreement with the recommendations of the SDC. For the PDS data, we assumed the fixed normalization with respect to the MECS of 0.86. 

\section{The model}
\label{s:model}

\begin{figure}
\centerline{\includegraphics[width=78mm]{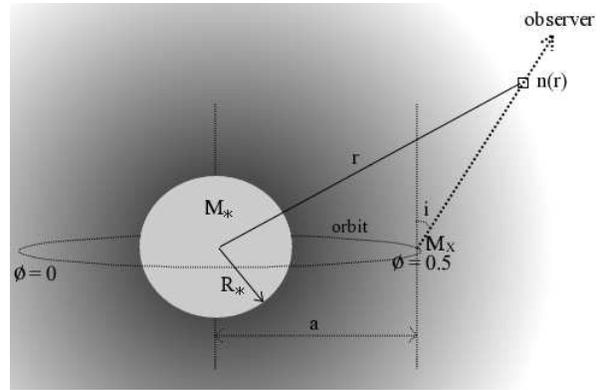}}
\caption{A schematic representation of the geometry of Cyg X-3. The
spectroscopic inferior conjunction corresponds to the orbital phase of $\phi =
0.5$. The trajectory of the compact object is plotted in the rest frame of the
primary. The shaded area represents a spherically symmetric, homogeneous,
stellar wind, with the darkness increasing with the density, dependent on the
distance from the centre of the primary, $r$.} 
\label{binary}
\end{figure}

Spectra observed from Cyg X-3 are determined by the form of the intrinsic spectra and their modification by the intervening medium. The latter consists of absorption in the interstellar medium and absorption and re-emission in the local medium, which consists primarily of the stellar wind. The intrinsic spectrum and the physical state of the stellar wind are coupled through complex plasma processes. Namely, as the X-ray photons travel through the wind, they heat and ionize it, which affects both the wind absorption and emission. Thus, when we try to model the observed spectra, we encounter a complication that the observed spectrum depends on both the intrinsic spectrum and the distribution of the wind density and its physical state along the line of sight, which, in turn, depends on the shape and flux of the intrinsic spectrum. Also, the wind acceleration depends on the irradiating spectrum and the ionization state. Below, we deal with some of those issues using the broad-band \sax\/ data. We use detailed physical description of both the intrinsic continua and the wind absorption.  

\subsection{The binary}
\label{s:binary}

A schematic picture of the binary is shown in Fig.\ \ref{binary}. In calculating the distribution of the stellar wind, we assume for simplicity that the primary is at rest. We also assume that the orbit is circular. The mass and the stellar radius of the primary are denoted as $M_{\star}$ and $R_{\star}$, respectively. The mass of the compact object (either a neutron star or a black hole) is denoted as $M_{\rm X}$. 

The orbital phase $\phi = 0$ corresponds to spectroscopic superior conjunction
(i.e., the compact object being behind the star), and $\phi$ is within the 0--1
interval. The distance between the centres of mass, $a$, follows from the
Kepler law, 
\begin{eqnarray}
\lefteqn{a^3 = { G (M_{\rm X} + M_{\star}) \over 4 \upi^2 } P^2,}\\
\lefteqn{a\simeq 3.1\times 10^{11} \left(M\over 30 \msun\right)^{1/3}{\rm cm}\simeq 4.47 \left(M\over 30 \msun\right)^{1/3} \rsun,}
\label{kepler}
\end{eqnarray}
where $G$ is the gravitational constant and $M=M_{\rm X} + M_{\star}$.

\begin{figure}
\centerline{\includegraphics[width=70mm]{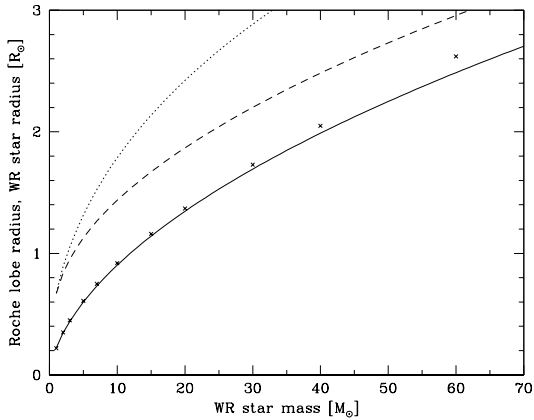}}
\caption{Comparison of the helium WR stellar radii with the Roche-lobe radii. The crosses give the model radii of Langer (1989), and the solid curve gives the approximation given by equation (20) in that paper. The dotted and dashed curves give the Roche lobe radii in Cyg X-3 calculated for the mass of the compact object of $1.4\msun$ and $20\msun$, respectively. 
} 
\label{Roche}
\end{figure}

The primary is a massive WR star, which emits a strong radiatively driven stellar wind. We compare the stellar radii, as given by Langer (1989), with the WR Roche lobe radii, calculated following Eggleton (1983) for the separation given by equation (\ref{kepler}). The radii given by Langer (1989) correspond to the star itself, without including the wind. Since the wind is, in general, optically thick, those radii are smaller than the photospheric radii, i.e., those corresponding to the optical depth of $\tau=2/3$. In Fig.\ \ref{Roche}, we plot the stellar radii corresponding to WR stars with the abundances dominated by He. As we see, the star always fits well within the Roche lobe, regardless of the compact object being a neutron star or a black hole. 

\subsection{The wind}
\label{s:wind}

We assume the wind to be smooth, spherically symmetric, non-rotating, stationary and radiatively driven. We follow Springmann (1994) and parameterize the wind velocity as follows, 
\begin{equation}
v(r) = v_{\star} + (v_{\infty}- v_0)\sqrt{\beta_1 x + (1 - \beta_1 ) x^{\beta_2}},
\label{velocitylaw}
\end{equation}
where $x = 1-R_{\star}/r$, $r$ is the distance from the centre of the primary,
$v_{\infty}$ is the wind terminal velocity, $\beta_1$ and $\beta_2$ are
constants which parametrize the wind acceleration rate, and $v_0$ is the
initial wind velocity, which we set to a value of 20 km s$^{-1}$, approximately corresponding to the atmosphere sound speed. According to Springmann (1994), equation (\ref{velocitylaw}) better describes winds of some WR stars than the usual wind velocity law of $v_{\infty} (1 - R_{\star}/r)^{\beta}$.

\begin{figure}
\centerline{\includegraphics[width=72mm]{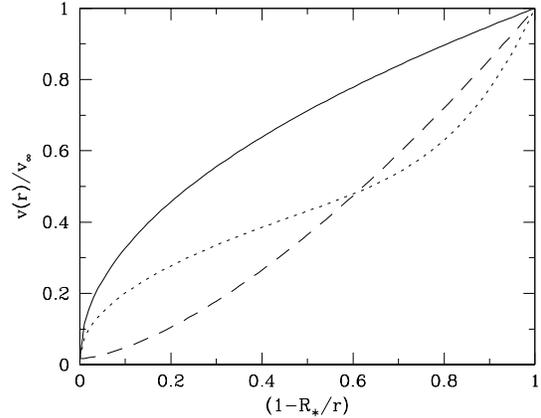}}
\caption{Wind velocity laws for different values of $\beta_1$ and $\beta_2$.
The solid curve represents a fast accelerating wind with $\beta_1=0,\,
\beta_2=1$, the dashed curve with $\beta_1=0,\, \beta_2=3$ is for a slowly
accelerating wind, and the dotted curve with $\beta_1=0.35,\, \beta_2=8$
corresponds to winds of some WR stars (Springmann 1994).} 
\label{velocities}
\end{figure}

\begin{table*}
\centering
\caption{The model parameters and the mass fractions and abundances relative to H for the most abundant elements (from Hamann \& Gr\"afener 2004)
for early-type, WNE, and late-type, WNL, WR stars. The H mass fraction in WNE
stars is close to zero, here we set it to an arbitrarily chosen small value.
}
\begin{tabular}{lllll}
\hline
parameter       &\multicolumn{2}{c}{WNE} & \multicolumn{2}{c}{WNL}            \\
\hline
$\mu$ & \multicolumn{2}{c}{1.33}  & \multicolumn{2}{c}{1.00} \\

$\mu_0$ & \multicolumn{2}{c}{4.05}  & \multicolumn{2}{c}{2.52} \\

$v_{\infty}$ [cm s$^{-1}$] & \multicolumn{2}{c}{$1.6 \times 10^8$} & \multicolumn{2}{c}{1.3 $\times 10^8$}\\                       
$A_{\rm Fe}$ & \multicolumn{2}{c}{1.78} &  \multicolumn{2}{c}{1.48} \\
$A_{>{\rm He}}   $ & \multicolumn{2}{c}{1.89} &  \multicolumn{2}{c}{1.58} \\
\hline
element & mass fraction     & $n_{\rm element/H}$  & mass fraction & $n_{\rm element/H}$\\ 
\hline
H  & $1.0\times 10^{-5}$  & 1                  & 0.2                  & 1   \\
He & 0.98                 & 2.46 $\times 10^4$ & 0.78                  & 0.98 \\
C  & $1.0 \times 10^{-4}$ & 0.84               & $1.0 \times 10^{-4}$  & 4.2 $\times 10^{-5}$\\
N  & $1.5 \times 10^{-2}$ & 1.08 $\times 10^2$ & $1.5 \times 10^{-2}$  & 5.4 $\times 10^{-3}$\\
O  & $5.5\times 10^{-3}$  & 34.65              & $5.5\times 10^{-3}$   & $1.73\times 10^{-3}$\\
Ne & $1.42\times 10^{-3}$ & 7.07               & $1.42\times 10^{-3}$  & $3.54\times 10^{-4}$\\
Fe & $1.4 \times 10^{-3}$ & 2.53               & $1.4 \times 10^{-3}$  & 1.26$\times 10^{-4}$\\
\hline
\end{tabular}
\label{abund}
\end{table*}

We consider three different parameterizations of the velocity law, also plotted
in Fig.\ \ref{velocities},

\begin{enumerate}
\item $\beta_1 = 0$, $\beta_2 = 1$ -- a very fast accelerating wind;

\item $\beta_1 = 0$, $\beta_2 = 3$ -- a wind with a slow initial acceleration
and a fast one at large distances;

\item $\beta_1 = 0.35$, $\beta_2 = 8$ -- an observational parameterization of
some WR stellar winds, steep at the base, flat in the middle and again steeply
accelerating at large distances (Springmann 1994). 
\end{enumerate}

The ion number density (averaged over possible clumping of the wind, see in Section \ref{s:transfer} below) at $r$ is related to $v(r)$ and the mass-loss rate, $\dot M$, via the continuity equation, 
\begin{equation}
\langle n(r)\rangle = {\dot M \over 4 \upi m_{\rm H}\mu_0 r^2 v(r) },
\label{n_r}
\end{equation}
where $m_{\rm H}$ is the hydrogen mass and $\mu_0$ is the mean ion molecular weight. The wind density along the line of sight is then determined by $\mu_0$, $R_{\star}$, $\dot M$, $M$, $\phi$, and the inclination, $i$.

The elemental abundances of WR stars are significantly different than the
solar. Hydrogen, if present at all, is heavily depleted in favour of helium and
nitrogen. While the atmospheres of early-type nitrogen sequence (WNE) WR stars
are hydrogen free, the late-type (WNL) ones contain some residual H. Van
Kerkwijk et al.\ (1996) classified Cyg X-3 as the WN 4--8 type (i.e.,
intermediate between those two types) on the basis of the I and K-band emission
lines, but did not find any sign of hydrogen in the infrared spectra. Here, we
perform and compare calculations for two different chemical compositions
characteristic for WNE and WNL stars. The WR star abundances of H, He, C, N and Fe come from grid models of Hamann \& Gr\"afener (2004), and are listed in Table \ref{abund}. The remaining abundances were set to the solar values in the sense that the mass fractions are set to the solar ones while their relative abundances were scaled to the actual abundance of H.

For a given set of most abundant elements, the H average number density is
\begin{equation}
\langle n_{\rm H} (r)\rangle ={\langle n(r)\rangle \over 1 + n_{\rm He/H} + n_{\rm C/H} + n_{\rm N/H} + n_{\rm O/H} + n_{\rm Ne/H} + n_{\rm Fe/H} },
\end{equation}
where $n_{\rm He/H}$, $n_{\rm C/H}$, $n_{\rm N/H}$, $n_{\rm O/H}$, $n_{\rm
Ne/H}$, $n_{\rm Fe/H}$ represent the abundances of the respective elements
relative to hydrogen. We set the H mass fraction for a WNE star to a small
value instead of 0, which would otherwise prevent us from defining the
abundances with respect to H. The abundances used by us are listed in Table
\ref{abund}. WNE stars tend to have faster winds than WNL ones (Hamann,
Koesterke \& Wessolowski 1995). Therefore, we choose the velocity range of
$v_{\infty} = 1300$--1600 km s$^{-1}$ (van Kerkwijk et al.\ 1996), and
attribute the lower limit to WNL stars and upper limit to WNE stars. The wind
is likely to be not smooth, but clumped, which effect we include in the way
described in Section \ref{s:transfer} below. 

\subsection{The intrinsic X-ray spectra}
\label{s:eqpair}

In spite of decades of research, the form of the intrinsic spectra of Cyg X-3 
remains highly uncertain. The reason for it is the very strong X-ray absorption 
in this source. As mentioned above, the state of the absorbing medium is coupled 
to the form of the irradiating spectrum. This introduces strong coupling between various fit parameters. The X-rays are further absorbed in the interstellar medium and reach the observer with almost no photons below 1 keV. Only the high energy part of the spectrum remains relatively unabsorbed. In order to efficiently deal with this problem, we start with a simple phenomenological description of the absorption while fitting the parameters of the intrinsic spectrum. 

The X-ray radiation, most likely, originates in the vicinity of the compact
object, and it is produced via accretion. The accretion disc emit soft blackbody
photons. Those photons are Comptonized in a hot plasma in the vicinity of the
disc. The plasma contains energetic electrons, which distribution may be
hybrid, i.e., containing both a Maxwellian part and a nonthermal tail beyond
it. Similarly to Vilhu et al.\ (2003), Hjalmarsdotter et al.\ (2004) and H08, we model the continuum emission of such a plasma with the {\tt eqpair}
model (Coppi 1999; see also Gierli\'nski et al.\ 1999), as implemented in the {\tt xspec} spectral fitting software (Arnaud 1996). The model calculates the spectrum from a spherical plasma. The distributions of photons and electrons are assumed to be isotropic and homogeneous. Some major model parameters are defined via the compactness,
\begin{equation}
\ell \equiv {L \sigma_{\rm T} \over R m_{\rm e} c^3},
\label{ls}
\end{equation}
where $L$ is a luminosity, $R$ is the characteristic size of the plasma, and $\sigma_{\rm T}$ is the Thomson cross section. In particular, we define the compactness, $\ell_{\rm s}$, corresponding to the soft photons irradiating the hot plasma with the luminosity, $L_{\rm s}$. The soft photon spectrum is taken here to be a disc blackbody ({\tt diskpn} of Gierli\'nski et al.\ 1999), which is characterized by the maximum blackbody temperature, $T_{\rm bb}$. The normalization, $N_{\rm C}$, of the {\tt eqpair} model equals that of the input disc emission, {\tt diskpn} (Gierli\'nski et al.\ 1999).

The electron distribution contains a Maxwellian part, with the optical depth corresponding to ionization electrons, $\tau_{\rm p}$. The thermal electrons are heated, i.e., supplied with the power, $L_{\rm th}$. If the plasma contains also a nonthermal part, it is assumed that the electrons are accelerated at a power-law rate $\propto \gamma^{-\Gamma_{\rm inj}}$, where $\gamma$ is the Lorentz factor and $\Gamma_{\rm inj}$ is the acceleration power-law index. The electrons are accelerated out of the thermal distribution, but in steady state the same number of electrons returns to it after losing their energy. The total power in the acceleration is $L_{\rm nth}$, and the total power supplied to the plasma is $L_{\rm h}= L_{\rm th} + L_{\rm nth}$. We then define three further compactness parameters, $\ell_{\rm th}$, $\ell_{\rm nth}$, and $\ell_{\rm h}$, according to equation (\ref{ls}). The physical processes taken into account are nonthermal electron energy losses via Comptonization and Coulomb losses with the thermal plasma, cooling and heating of the thermal plasma by Comptonization, and production of $e^\pm$ pairs. A process also taken into account, but usually negligible in luminous accretion flows, is bremsstrahlung. Then, the steady-state electron distribution, including the temperature of the thermal part, $T_{\rm e}$, is solved for self-consistently, and the corresponding emitted photon spectrum is calculated. The total steady-state optical depth may be higher than $\tau_{\rm p}$ due to $e^\pm$ pair production. The overall hardness of the spectrum is determined by the ratio, $\ell_{\rm h}/\ell_{\rm s}$, i.e., the amplification factor of the Comptonization.

A fraction of the X-rays is scattered back to the optically-thick medium
emitting blackbody photons (presumably the accretion disc), and it is reflected with an energy-dependent albedo (Magdziarz \& Zdziarski 1995). The relative strength of reflection is parametrized by $\Omega/2 \upi$, where $\Omega$ is the
effective solid angle subtended by the reflector. If the primary emission is
isotropic and neither the primary emission nor the reflected one is
obscured or attenuated (apart from any external absorption, e.g., in the wind
and the interstellar medium), $\Omega$ is also the actual solid angle subtended
by the reflector. However, the intrinsic X-ray source may be obscured, e.g., by
an inner torus-like structure, whereas the reflection from the torus back wall
can be seen directly. Then, $\Omega/2 \upi$ is a measure of the degree of
obscuration of the central source. The reflecting surface is allowed to be
ionized, with the ionization parameter defined as $\xi\equiv L_{\rm ion}/n_{\rm refl} R_{\rm refl}^2$, where $L_{\rm ion}$ is the 5 eV--20 keV luminosity and $n_{\rm refl}$, $T_{\rm refl}$ is the density and the temperature, respectively, of the reflector located at a distance, $R_{\rm refl}$, from the illuminating source. We use here the ionization model by Done et al.\ (1992), which although is not applicable to highly ionized Fe (Ballantyne, Ross \& Fabian 2001), is sufficiently accurate at low ionization, such as that found for our data below. We hereafter denote the total intrinsic continuum, which includes the disc blackbody emission and both Compton upscattering and Compton reflection, as $F_{\rm C}$.

The primary and reflected emission is then absorbed in the intervening media.
One of the absorbers is the interstellar medium, with the line-of-sight column
density, $N_{\rm H0}$. It has been estimated, in units of $10^{22}\,{\rm cm}^{-2}$, as 1.7, 1.4 and 1.42 by Chu \& Bieging (1973), Lauqu\'e, Lequeux \& Nguyen-Quang-Rieu (1972), and Dickey \& Lockman (1990), respectively. Here, we assume $N_{\rm H0} = 1.5 \times 10^{22}\, {\rm cm}^{-2}$. We assume that absorbing medium is neutral, model it with the {\tt wabs} model
(Morrison and McCammon 1983), and denote the resulting energy-dependent
attenuation factor as $A_0(E)$, where $E$ is the photon energy.

In addition, there is a strong local absorption, apparently in the stellar wind, as none of our data can be satisfactorily fitted with interstellar
absorption only. However, as discussed above, the intrinsic continuum and the
absorbing wind parameters are strongly mutually dependent. Thus, we start with
a simple model of intrinsic absorption modifying the {\tt eqpair} continuum.
Namely, we assume a product of complete and partial covering, in both cases
assuming neutral absorbing medium. The column densities are $N_{\rm H,tot}$ and
$N_{\rm H,part}$, respectively, and the second absorber covers a fraction, $C$,
of the source,
\begin{equation}
A_{\rm par}(E) = \left[C {\rm e}^{-\sigma(E) N_{\rm H, par}}+ 1 - C
\right]{\rm e}^{-\sigma(E) N_{\rm H, tot}},
\label{partial}
\end{equation}
where $\sigma(E)$ is the absorption cross section. This phenomenological model
can fit the data relatively well, and was used for Cyg X-3 spectra by Vilhu et
al.\ (2003), Hjalmarsdotter et al.\ (2004) and H08.

Furthermore, the data show a relatively strong line related to the Fe K
complex. We model it as a single Gaussian with the centroid energy, $E_{\rm
Fe}$, the width, $\sigma_{\rm Fe}$ and the normalization, $I_{\rm Fe}$. The iron lines originate both from reflection and the local absorber. We denote this component as $F_{\rm Fe}(E)$.  

The total model has then the form,
\begin{equation}
F(E) = A_0(E) A_{\rm par}(E) \left[ F_{\rm C}(E) + F_{\rm Fe}(E) \right].
\label{phenomenon}
\end{equation}
Its main free parameters are the Comptonization amplification factor, $\ell_{\rm h}/\ell_{\rm s}$, the non-thermal fraction, $\ell_{\rm nth}/\ell_{\rm h}$, $\tau_{\rm p}$, $kT_{\rm bb}$, the reflection strength, $\Omega/2 \upi$, $\xi$, $\Gamma_{\rm inj}$, $N_{\rm H,tot}$ and $N_{\rm H,par}$, and $C$. Then, we assume fixed values of $\ell_{\rm s} = 100$, $T_{\rm refl} = 10^7$K (important for reflection only), the characteristic size of the plasma (important only for the Coulomb and bremsstrahlung processes), $R = 10^7$ cm, the minimum and maximum Lorentz factors the accelerated electrons, $\gamma_{\rm min} = 1.3$, $\gamma_{\rm max} = 1000$, respectively, and the inclination angle, $i = 60\degr$ (important for the disc emission and reflection). We neglect relativistic broadening of the reflected spectrum and the Fe K line.

The relative iron abundance, $A_{\rm Fe}$, and the relative abundance of
elements heavier than He, $A_{\rm >He}$, used by {\tt eqpair}, assume the
solar elemental abundances of Anders \& Ebihara (1982). They are important for
reflection, and determine the relative importance of bound-free absorption
(determined in the X-ray regime mostly by elements heavier then He) to
scattering (determined by the electron abundance). We rescale them with respect
to the actual electron abundance in the wind with the WR-star abundances, see
Table \ref{abund}. Namely,
\begin{equation}
A_{\rm Fe} = {(n_{\rm Fe}/n_{\rm e})_{\rm WR} \over (n_{\rm Fe}/n_{\rm e})_{\odot}},\qquad A_{\rm >He} = {(n_{\rm >He}/n_{\rm e})_{\rm WR} \over (n_{\rm >He}/n_{\rm e})_{\odot}},
\end{equation}
where the WR and solar subscripts refer to the WR-star abundances used by us and to those of Anders \& Ebihara (1982), respectively, and, for each, the ratios to $n_{\rm e}$ are given by,
\begin{eqnarray}
\lefteqn{{n_{\rm Fe} \over n_{\rm e}} = {n_{\rm Fe/H} \left(\mu_0/\mu-1
\right)^{-1} \over  1 + n_{\rm He/H} + n_{\rm C/H} + n_{\rm N/H} + n_{\rm O/H} + n_{\rm Ne/H} + n_{\rm Fe/H} },}\\
\lefteqn{
{n_{\rm >He} \over n_{\rm e}} = {(n_{\rm C/H} + n_{\rm N/H} + n_{\rm O/H} + n_{\rm Ne/H} + n_{\rm Fe/H}) \left(\mu_0/\mu-1
\right)^{-1} \over  1 + n_{\rm He/H} + n_{\rm C/H} + n_{\rm N/H} + n_{\rm O/H} + n_{\rm Ne/H} + n_{\rm Fe/H} },}
\end{eqnarray}
where $\mu$ is the mean molecular weight.

\subsection{Radiative transfer in the wind}
\label{s:transfer}

The X-ray spectrum emitted by the hot plasma is then transferred in the stellar
wind. This radiative transfer is modelled using the {\tt cloudy} photoionization
code (v.\ 6.02, see Ferland et al.\ 1998). The main input parameters of {\tt
cloudy} are the shape of the irradiating continuum, the density profile along
the line of sight, and the wind elemental abundances. The code also allows to
account for the effect of possible wind clumping, with clouds filling a
fraction, $f$, of the volume, and with empty region between the clumps.

For a given $\dot M$, the density, $n$, of the matter inside the clumps
equals $\langle n\rangle/f$, i.e., it is $1/f$ times higher than that of a smooth wind, $\langle n\rangle$. Although neither $\langle n\rangle$ nor the column density change, the state of the plasma does. In photoionized plasmas, the volume-averaged ionization rate is $\propto (\langle n\rangle/f)f=\langle n\rangle$. On the other hand, the recombination rate is proportional to $n^2$, and its volume-averaged rate is $\propto (\langle n\rangle/f)^2 f=\langle n\rangle^2/f$, i.e., it increases. Thus, for a fixed radiation field, increasing the density decreases the average effective charge of a given element, i.e., the plasma becomes less ionized. 

In our calculations, we also consider a two-phase wind, in which the region between the clumps is not empty. We approximate that situation by considering two {\tt cloudy} models, one with a smooth, $f=1$, optically thin wind,
and one with the clumps, $f\ll 1$. The results of the two calculations
are then superimposed on the intrinsic spectrum, see Section \ref{s:cloudy}
below.

We note that in our calculations we neglect the effect of X-ray irradiation on the wind velocity and density. Such calculations cannot be done within the framework of our model. They would require iterative hydrodynamic calculations of the entire wind structure, not only the effect of the irradiation on the line of sight. Since we neglect the effect of the X-ray radiation force on the wind, we consequently do not consider the pressure balance in the accelerated wind, which would taking into account the pressures of radiation and each phase of the wind as well as of the wind inertia. We also point out that {\tt cloudy} does not allow for the filling factor with an arbitrary dependence on position, and thus it would not allow us to achieve a global pressure balance of the two wind phases. 

We use the wind density distribution along the line of sight from the X-ray 
source calculated for a wind spherically symmetric with respect to the primary 
star. This density distribution is then used by {\tt cloudy} to create a cloud 
which is spherically symmetric with respect to the X-ray source, an 
approximation required by that code. Also, all of the radiation scattered from 
(or reemitted by) one side of the wind back to the opposite side is included in
calculations of the ionization structure and line emission (which is called the
closed geometry in {\tt cloudy}). It is also important to note that in the case of closed geometry, {\tt cloudy} does not calculate the effects of electron
scattering, which may be important for electron column densities substantially greater than $10^{23}$ cm$^{-2}$. 

For the irradiating continuum, we use the results of the first stage of the calculations assuming the phenomenological absorption, equations 
(\ref{partial}--\ref{phenomenon}). We create a set of table models in the {\tt xspec} format (both additive and multiplicative) for the HS and SS spectra and for a number of the wind parameters, which we then use for the final spectral fitting. 

\begin{figure}
\centerline{\includegraphics[width=70mm]{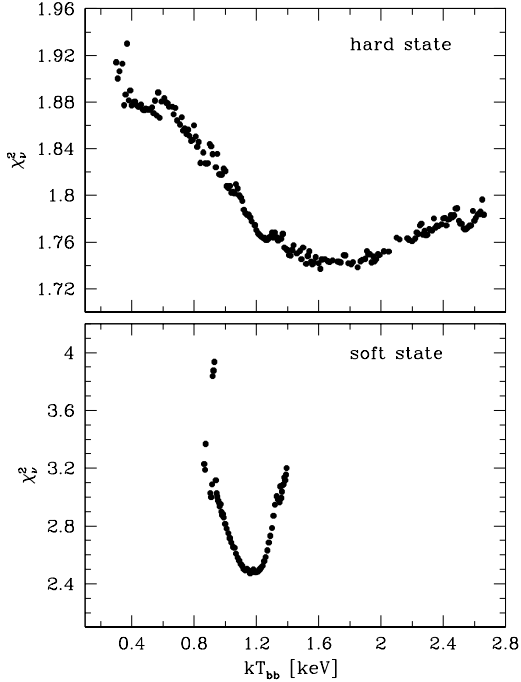}}
\caption{The dependencies of $\chi^2_{\nu}$ on $kT_{\rm bb}$ in the HS and SS
for the phenomenological treatment of absorption and the WNE stellar type (see
Section \ref{s:phenomenon}). The number of degrees of freedom is 466 and 521 in
the HS and SS, respectively. } 
\label{fits}
\end{figure}

\begin{figure}
\centerline{\includegraphics[width=83mm]{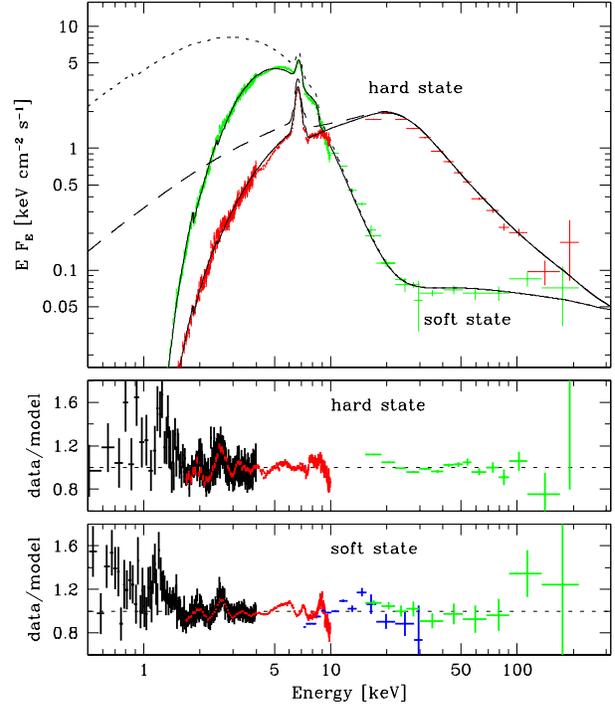}}
\caption{The top panel shows the best-fit unfolded spectra (crosses) and the best-fit models (solid curves) obtained with the phenomenological treatment of absorption, see Section \ref{s:phenomenon}). The dashed and dotted curves show the unabsorbed spectra in the HS and SS, respectively. The two lower panels show the corresponding data to model ratios. The black, red, blue and green symbols correspond to the LECS, MECS, HPGSPC, and PDS data, respectively.
} \label{eeuf}
\end{figure}

One issue we need to consider here is the effect of the blackbody photons
emitted by the primary on the wind. Those UV photons are important for the
balance between Compton heating and cooling. In the limit of complete
ionization, the gas reaches the Compton temperature, $T_{\rm C}$,
\begin{equation}
T_{\rm C} = {\int E F(E) {\rm d}E \over 4 k \int F(E) {\rm d}E},
\label{comptont}
\end{equation}
(e.g., Begelman, McKee \& Shields 1983), where the irradiating flux, $F$, is
the sum of the X-ray flux, $F_{\rm X}$, and the stellar emission, $F_\star$. The numerator and denominator in equation (\ref{comptont}) correspond to the heating and cooling, respectively. We note that for two separate sources of illumination, the ratio between the two fluxes is a strong function of the position in the wind, except for large distances from the binary, where, in the optically thin case, it reaches a constant. That approximation, with the sum of the X-ray and UV sources as ionizing the wind, was appropriate for absorption far away from the binary plane considered by Szostek \& Zdziarski (2007) for Cyg X-1. However, it cannot be applied to wind absorption close to the binary plane, as it is the case in our present study.

A luminous WR primary may be a significant contributor to the cooling. On the other hand, the X-ray source with $L_{\rm X}\sim 10^{38}$ erg s$^{-1}$ (Vilhu et al.\ 2003) dominates the heating except very close to the primary. However, the {\tt cloudy} code does not allow to 
include two independent and spatially resolved illuminating sources. Therefore, 
we consider the X-rays as the only source of the wind illumination. Since we 
neglect the existence of the stellar cooling, the model gives an upper limit on 
the wind temperature. Thus, it is somewhat more ionized, and less absorbing, in our model than in reality. Hence, it gives upper limits on the wind density and $\dot M$ for a given set of model parameters. In Section \ref{cooling} below, we estimate the maximum effect of the wind illumination by the companion on our final results, and find it very minor.

\section{Results}
\label{results}

\subsection{The phenomenological treatment of absorption}
\label{s:phenomenon}

We first fit the data using the phenomenological treatment of absorption as a 
product of complete and partial covering, equation (\ref{phenomenon}). Still, 
even this simplified model shows a complex $\chi^2$ structure, with a number of 
local minima and strong correlations between its parameters. In particular, the 
values of the maximum blackbody temperature, $T_{\rm bb}$, and of the absorber 
column densities are strongly correlated. The absorption removes almost all 
photons below 1 keV, which results in very weak constraints on the form of the 
intrinsic spectrum in that range. For very similar fits to the data, those 
intrinsic spectra can have very different form. This uncertainty would also 
affect the ionization structure of the wind (considered in Section 
\ref{s:cloudy} below).

Thus, we first perform a detailed search for the global $\chi^2$ minimum as a 
function of $T_{\rm bb}$. Since the HPGSPC and PDS spectra are only weakly 
affected by the absorption whereas the lower energy spectra are strongly 
affected, we temporarily increase the relative weight of the HPGSPC and PDS data 
by decreasing their measurement errors by a factor of 10. Also, given the 
relatively poor quality of the high energy data in the SS, we use for it the 
constant $\Gamma_{\rm inj}=2$. Also, the Fe K line complex is relatively weak in 
the SS, and we thus fix $\sigma_{\rm Fe}=0.2$ keV. We also find that changing 
the abundances between the WNE and WNL models (Table \ref{abund}) has little 
effect on the intrinsic spectra. Thus, we consider here only the WNE model. 

Because the data are averaged over $\sim$5 orbital periods, it is unlikely that all discrete spectral features would be fully correctly reproduced with
a single-phase model. Therefore, we add a 3 per cent systematic error to the model. Because of a large number of local minima, the default {\tt xspec}
minimization method of modified Levenberg-Marquardt based on the {\tt curfit}
routine from Bevington \& Robinson (1992) was found to be inefficient. Thus, we use the {\tt migrad} method based on the CERN {\tt minuit} code (James \& Winkler 2004), which is also implemented in {\tt xspec}. The latter method is more time consuming but was found to give much better results than the former.

We present the resulting dependencies of $\chi^2_\nu$ on $kT_{\rm bb}$ in Fig.\
\ref{fits}. We see a relatively narrow minimum in the SS at $\sim$1.2 keV. On
the other hand, the minimum in the HS is much broader, at $\sim$1.5--1.9 
keV. Hereafter, we assume constant values of $kT_{\rm bb}=1.2$
keV and 1.7 keV in the SS and HS, respectively. The higher value of $kT_{\rm bb}$ in the HS may be due to the presence of some kind of a soft excess in that state, commonly found in black-hole binaries (e.g., Gierli\'nski \& Zdziarski 2005), or may be due to the origin of at least some of the soft photons from another form of matter, e.g., dense blobs in the wind (see Section \ref{s:cloudy} below). The HS and SS unfolded spectra, including the unabsorbed, intrinsic, spectra are shown in Fig.\ \ref{eeuf} (after restoring the original HPGSPC and PDS errors). The HS and SS spectra studied here are very similar to the spectra 1 and 5, respectively, obtained from averaging pointing observations of {\it Rossi X-Ray Timing Explorer\/} by Szostek \& Zdziarski (2004). 

The rather high minimum values of $\chi^2_{\nu}$ in both states point to the
model being too simplified. Its main problems can be identified by looking at the fit residuals in Fig.\ \ref{eeuf} (with the original HPGSPC and PDS
errors). Strong residuals at energies $\la 3$ keV are apparently from
neglecting the wind emission lines. Also, the Fe K complex is not accurately
treated by a single Gaussian. Still, the overall continuum shapes have been
found to be sufficiently accurate to calculate the effect of the wind
irradiation, see below.

\begin{figure}
\centerline{\includegraphics[width=83mm]{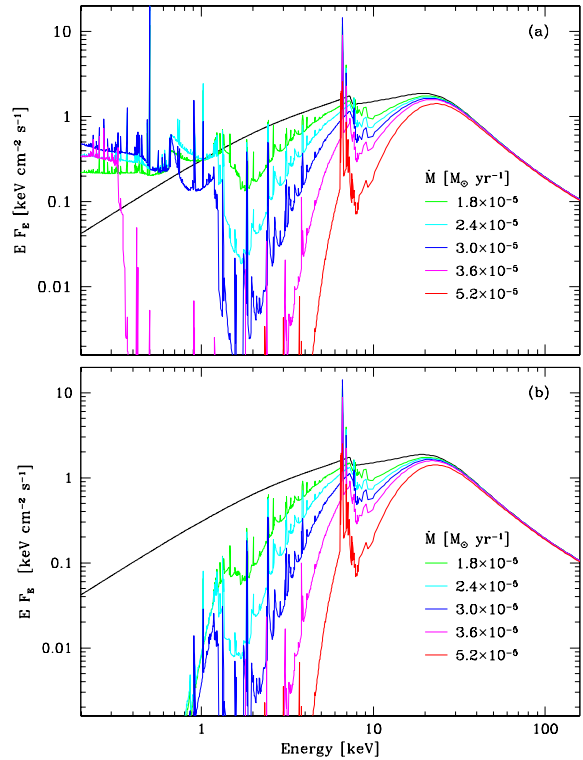}}
\caption{Example model spectra transmitted through the stellar wind for a range
of values of $\dot M$ (colour curves) and the rest of the parameters kept at
the values given in Section \ref{s:cloudy}. The original unabsorbed continuum
is shown by the black curve. (a) The spectra transmitted only through the
stellar wind. (b) The spectra transmitted through both the wind and the
interstellar medium in the direction of Cyg X-3. } 
\label{massloss}
\end{figure}

\subsection{Wind absorption and emission}
\label{s:cloudy}

We use the above best-fit unabsorbed models as the input spectra for the {\tt
cloudy} calculations of the radiative transfer in the stellar wind. The
13.6--$2\times 10^5$ eV ionizing luminosities (used by {\tt cloudy}) are $8.0 \times 10^{37}$ erg s$^{-1}$ and $2.8 \times 10^{38}$ erg s$^{-1}$ in the HS and SS, respectively.

Before the actual fitting, we calculate the effect of changing the wind 
parameters on the transmitted spectra. Since the fitted spectra are 
phase-averaged, we use the orbital phase in the middle between the superior and 
inferior conjunctions, $\phi=0.25$. This choice of $\phi$ has also the advantage 
that, assuming spherical symmetry, there is no dependence of the wind absorption 
on the inclination. (The only effect of $i$ is then on the disc blackbody and 
reflection components of the intrinsic continuum.) For these calculations, we 
consider the HS only but the conclusions also apply to SS. Then, we assume as 
the test default parameters the WNE stellar type, the velocity law (i) (with 
$\beta_1=0$, $\beta_2=1$), $R_{\star} = 1 \times 10^{11}$cm, the total mass of 
the two components, $M = 50\msun$, $f = 1$ and $\dot M = 2.0 \times 
10^{-5}\msun$ yr$^{-1}$. For these parameters, the column density of (mostly 
fully ionized) He along the line of sight is $N_{\rm He} =4.2 \times 10^{23}$ 
cm$^{-2}$. The corresponding Thomson optical depth is $\tau_{\rm w} \simeq 0.6$. 
We change one of the default parameters while keeping the rest at the above 
values. 

\begin{figure*}
\centerline{\includegraphics[width=125mm]{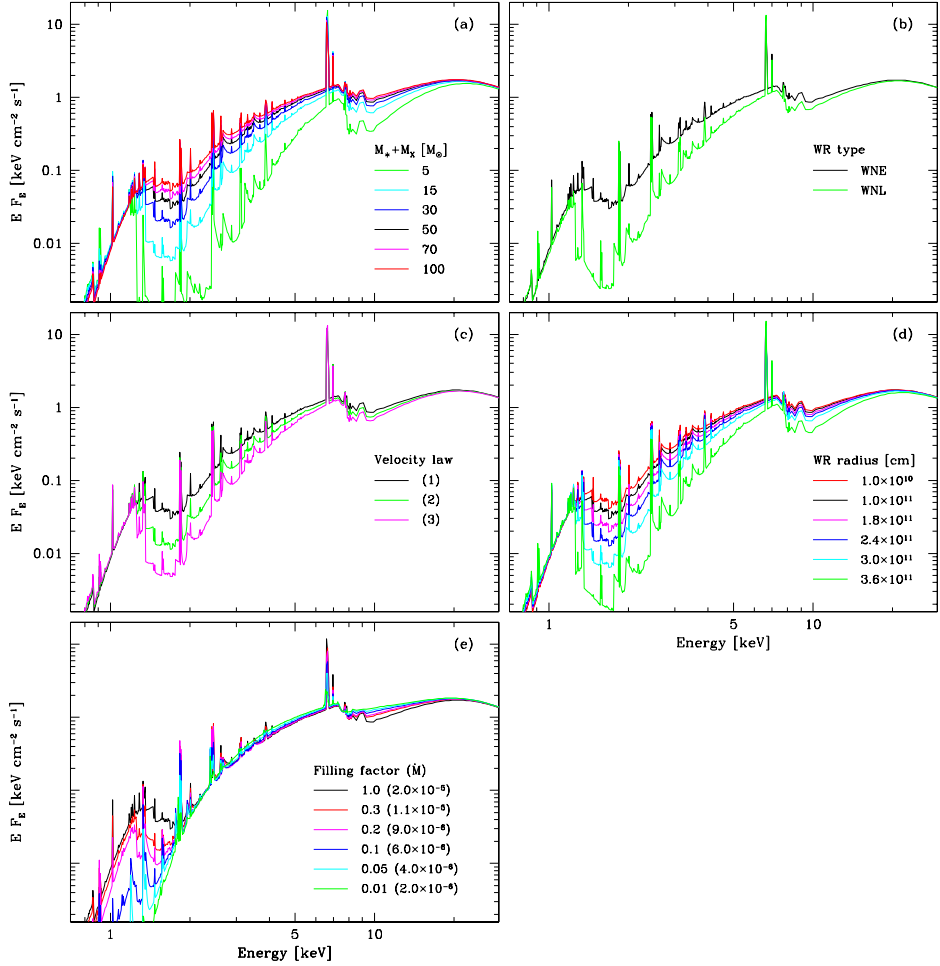}}
\caption{Example model spectra transmitted through both the wind and the 
interstellar medium. In each of the (a--e) panels, one parameter is varied
(identified by its colour) while the rest are kept at the initial values, given
in Section \ref{s:cloudy}. The black curves correspond to the initial value of
the variable parameter. (f) Coordinated variability of $f$ and $\dot M$. }
\label{tests}
\end{figure*}

Fig.\ \ref{massloss}(a) shows the spectra transmitted through the wind for 
$\dot M= (1.6$--$5.0) \times 10^{-5}\msun$ yr$^{-1}$. As
expected, the increasing $\dot M$ results in the increasing absorption. The wind
is strongly ionized and, except for the two highest values of $\dot M$, there
is almost no absorption at energies $\la 1$ keV. On the other hand, there is
relatively strong wind emission at those energies. Fig. \ref{massloss}(b) shows
the same spectra but transmitted also through the interstellar medium with
$N_{\rm H}$ of Cyg X-3. Now, all the emission at $\la 1$ keV is strongly
absorbed. We would like to draw the attention to the shape of the spectrum for
$\dot M = 2.4\times 10^{-5}\msun$ yr$^{-1}$. The elements which
contribute to absorption below 1 keV are completely ionized and the low energy
photons are not absorbed. The elements responsible for absorption above 1 keV
are only partially ionized and create a deep, broad absorption feature. The
combined effect of low absorption in the wind at low energies and strong
interstellar absorption creates a spectral shape with a strong soft excess at
$\la 2$ keV.

Fig.\ \ref{tests} shows the effect of changing the other parameters on
the spectra absorbed by both the wind and the interstellar medium. The wind of
the WNL-type star absorbs much more than that of the WNE type due to the lower
wind velocity of the former, Fig.\ \ref{tests}(b). Absorption decreases with
the increasing total mass (determining the separation) and the stellar radius
(which is due to the scaling of the wind velocity with $r/R_\star$, see Section
\ref{s:wind}), see Figs.\ \ref{tests}(a, d). Then the wind velocity laws
(i)--(iii) yield correspondingly increasing absorption, Fig.\ \ref{tests}(c).
In most cases, the effect on the spectra at $\sim 1.5$ keV is only minor, as
the elements responsible for absorption in that range are almost fully ionized
in the wind. 

Fig.\ \ref{tests}(e) shows the case where a coordinated decrease of both
$f$ and $\dot M$ leads to the overall form of the spectrum at $\ga 2$ keV
changing only slightly. On the other hand, the reduction of $f$ leads to a
decline and disappearance of the emission excess at $\la 2$ keV. The lower the
$f$, the higher the density (of the blobs), and the weaker the ionization,
reducing then the soft X-ray emission. Another effect we see is the change in
the spectral shape in the $\sim\! (6$--10) keV region. Here, the hotter, more
ionized, gas (at a higher $f$) produces stronger Fe emission and absorption
than the cooler one at lower $f$. 

\begin{figure*}
\centerline{\includegraphics[width=125mm]{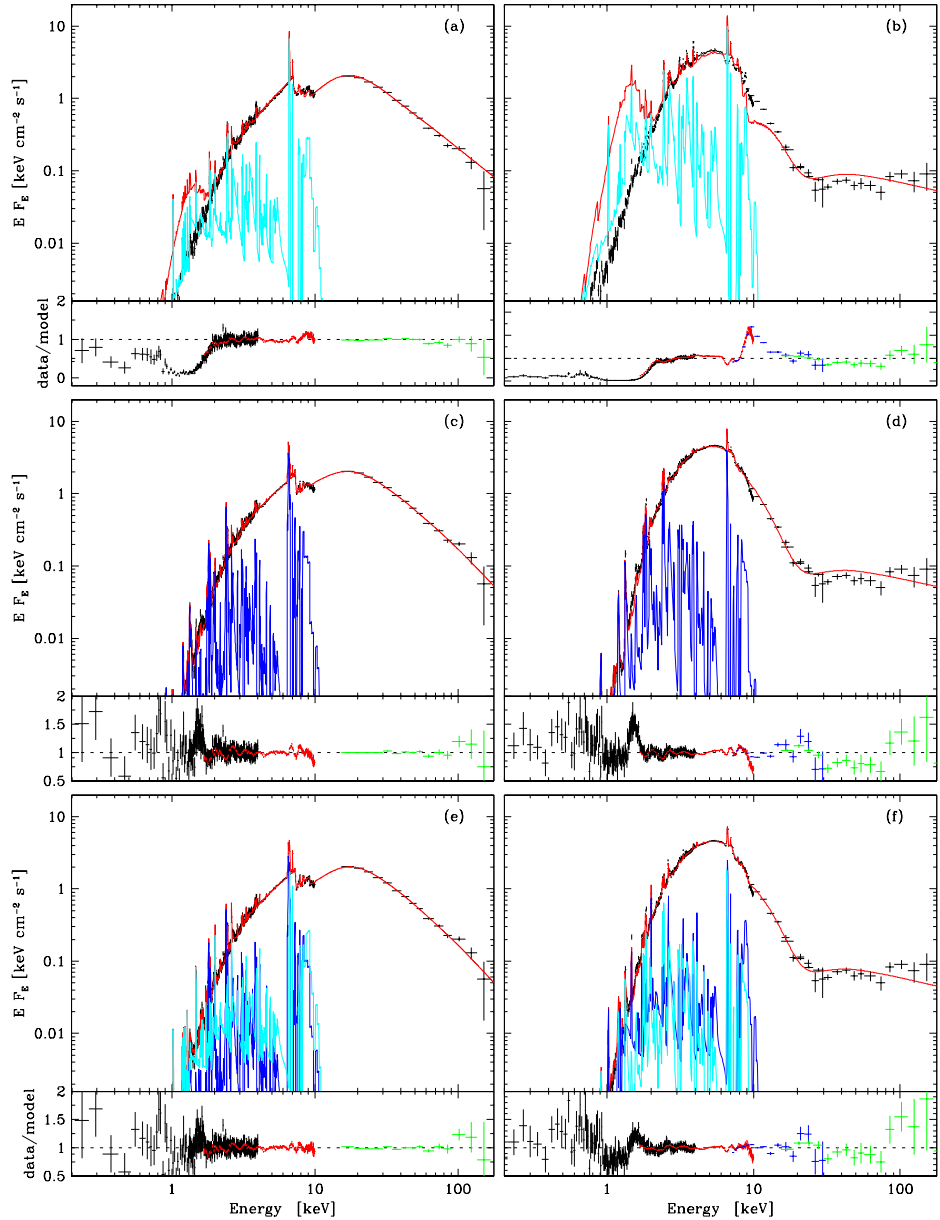}}
\caption{
The best-fit unfolded spectra and residuals fitted including wind
absorption and emission (for the case maximizing $\dot M$). The colours of the residuals are the same as in Fig.\ \ref{eeuf}. The HS fitted by the model of equation (\ref{model1}), (\ref{model2}), and (\ref{model3}) are shown in (a), (c) and (e), respectively. The SS fitted by the model of equation (\ref{model1}), (\ref{model2}), and (\ref{model3}) are shown in (b), (d) and (f), respectively. The emission of the hot wind, $f=1$, and the cool one, $f\ll 1$, are shown by the cyan and blue lines, respectively. 
}
\label{fits2}
\end{figure*}

\subsection{The minimum and maximum mass-loss rates}
\label{minmax}

Then, we would like to obtain the lower and upper limits on $\dot M$, which correspond to the model parameters maximizing and minmizing (for a given $\dot M$), respectively, the absorption. The minimum corresponds to the WNL-type star, the (iii) velocity law, and the smallest possible separation between the system
components. For the last requirement, we would like to consider the smallest
possible masses and the highest stellar radius. We assume $M_{\rm X} =
1.4\msun$, $M_{\star} = 5\msun$ (which is close to the smallest possible WR-star mass, and agrees with the estimates of Hanson, Still \& Fender 2000), which corresponds to $R_{\star} = 0.6 \rsun$ (Fig.\ \ref{Roche}). 

Next, the maximum of $\dot M$ corresponds to the WNE-type star, the (i)
velocity law, and the largest possible separation between the system
components. For the last requirement, we would like to consider the largest
possible masses and the lowest stellar radius. We assume $M_{\rm X} = 20\msun$, $M_{\star} = 50\msun$ (which is close to the largest possible WR-star mass using the standard models of Langer 1989), which corresponds to $R_{\star} = 2.3\rsun$. (We note that the assumed black hole mass is larger than that allowed by the constraints of Hanson et al.\ 2000).

For both the minimum and maximum we consider a number of values of $f$. For each
$f$, we create {\tt xspec} tables for a range of values of $\dot M$. We will
denote the resulting wind absorption as $A_{\rm w}(E)$, and the corresponding
line emission, as $F_{\rm w}(E)$. The following considerations apply to both minimum and maximum limit on $\dot M$, but we plot the results only for the maximum case. 

\subsection{Homogeneous hot wind}
\label{homogeneous}

We first consider the case with no clumps, $f = 1$. As before, we keep the
constant $\Gamma_{\rm inj}=2$ in the SS, and $kT_{\rm bb}=1.7$ keV
and 1.2 keV in the HS and SS, respectively. We also add a 3 per cent systematic error to the model (as in Section \ref{s:phenomenon}) and a 1.5 per cent systematic error to each data set. We drop other simplifying assumptions, in particular absorption is now only in the wind and the insterstallar medium. The structure of the model is,
\begin{equation}
F(E) = A_0(E) \left[ A_{\rm w,h}(E) F_{\rm C}(E) + F_{\rm w,h}(E) \right].
\label{model1}
\end{equation}
where the subscript `h' indicates that the wind with $f=1$ is hot. Below, we
also use the subscript `c' for the cooler gas in the compressed clumps with $f\ll 1$. In the spherical geometry with the central irradiating source used by {\tt cloudy}, the relative normalization of the emission-line component is unity. However, we use here {\tt cloudy} only as an approximation to the actual stellar wind geometry. Thus, we allow a free relative normalization of the emission-line component, but we find its best fit values to be $\sim$1. We denote that factor by $N_{\rm h}$, $N_{\rm c}$ for the cases with $f=1$, $f\ll 1$, respectively.

We show the resulting fits and their residuals in Figs.\ \ref{fits2}(a,b). We see the model provides rather bad fits, mostly due to its prediction of an excess emission at low energies. As explained in Section \ref{s:cloudy}, this excess is created by low absorption by the wind below $\sim$2 keV, see Fig.\ \ref{massloss}. The wind emission gives also some contribution, see the cyan lines in Figs.\ \ref{fits2}(a,b), but we can see it is rather minor (in comparison to the total model). We have tested this by fitting a model with the wind emission removed, and obtained rather similar soft excess in the model compared to the data at $\la$2 keV. Also, the SS model strongly overestimates the depth of the Fe {\sc xxv} and {\sc xxvi} edges at $\simeq$9 keV (see, e.g., Turner et al.\ 1992 for the Fe K edge and line energies). Thus, the data require the presence of additional denser material with lower ionisation. 

\subsection{Clumpy cool wind}
\label{clumpy}

In order to deal with the $\la 2$ keV residuals, we can reduce the filling
factor, $f$, see Fig.\ \ref{tests}(e). From our fitting, we find that the
maximum values of $f$ that remove those residuals are approximately $f=0.01$ and 0.005 in the HS and SS, respectively. The model is, 
\begin{equation}
F(E) = A_0(E) \left[ A_{\rm w,c}(E) F_{\rm C}(E) + F_{\rm w,c}(E) \right].
\label{model2}
\end{equation}

We show the resulting fits together with their residuals in Figs.\
\ref{fits2}(c, d). We see that this model provides a better fits than that with
$f=1$, and $\chi^2_{\nu}=1.98$ (470 d.o.f.) and 1.97 (524 d.o.f.) for the HS
and SS, respectively (for the case maximizing $\dot M$). Still, there are significant residua at various energies, in particular there are signs of highly ionized Fe in the data, which are not included in the model with a low $f$, in which the wind clumps are relatively weakly ionized. This indicates the presence of a hot phase. 

\subsection{Two phase wind}
\label{two_phase}

Indeed, it is highly unlikely that there is only vacuum in the interclump wind
medium. Therefore, we now model the wind as containing both the cool clumps and
a smooth hot gas between them. Because the two phases coexist together in the
same space, as discussed above, we account for it as a combination of two
separate {\tt cloudy} models calculated independently for the same system parameters and the illuminating source but for different $\dot M$ and $f$. This approximation is valid as long as the hot phase is optically thin, i.e., it does not substantially absorb soft X-rays but still emits in lines. The hot, very strongly ionized, phase may still cause some absorption around $\sim$9 keV. Each of the phases is now characterized by its own accretion rate. The model is now, 
\begin{equation}
F(E) = A_0(E) \left[ A_{\rm w,c}(E) A_{\rm w,h}(E) F_{\rm C}(E) + F_{\rm w,c}(E) + F_{\rm w,h}(E)\right].
\label{model3}
\end{equation}

\begin{figure}
\centerline{\includegraphics[width=77mm]{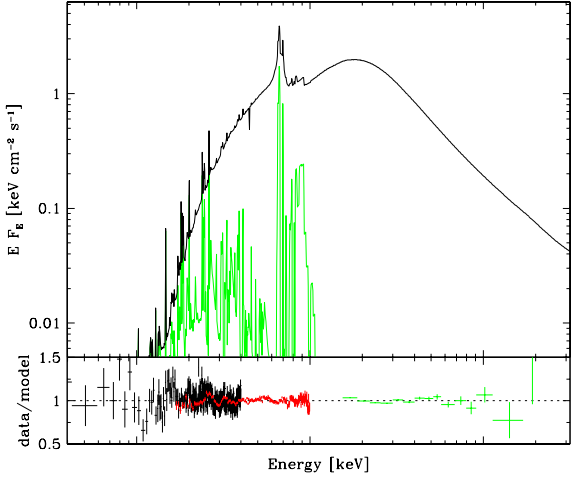}}
\caption{The best-fit absorbed HS model (for the case maximizing $\dot M$) including two-phase wind and an Fe K$\alpha$ line from Compton reflection (top solid curve). The wind emission is plotted in green below the top curve. The fit residuals are plotted in the same colours as in Fig.\ \ref{eeuf}. }
\label{bestHS}
\end{figure}

\begin{figure}
\centerline{\includegraphics[width=74mm]{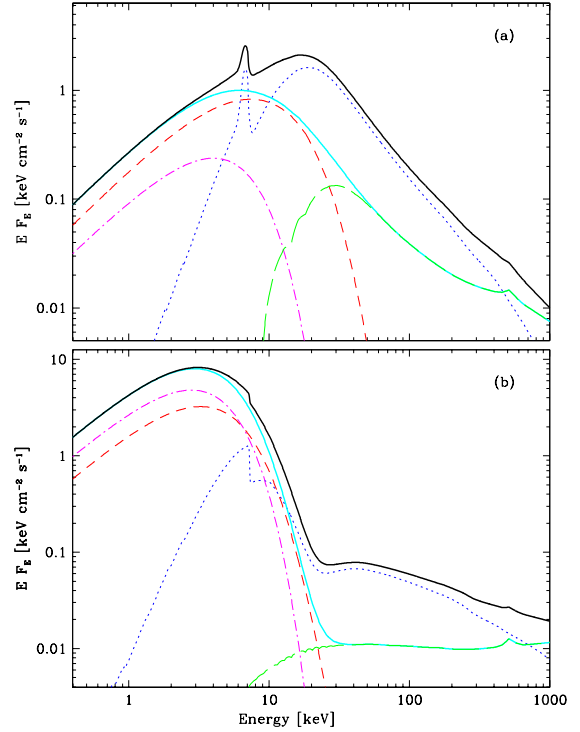}}
\caption{The unabsorbed spectra (solid curves) of the emission before the wind reprocessing shown together with their components for (a) the HS and (b) the SS. The unscattered disc blackbody, Compton upscattering by thermal and by nonthermal electrons, and the Compton reflection together with the Fe K$\alpha$ line are shown by magenta dot-dashes, red short dashes, green long dashes, and blue dots, respectively. The cyan solid curves show the intrinsic spectrum without the reflection component.}
\label{components}
\end{figure}

We show the resulting fits together with their residuals in Figs.\
\ref{fits2}(e,f). The overall fit quality is now very good, with 
$\chi^2_{\nu}=1.03$, 0.99 (468 d.o.f.) in the HS at the minimum and maximum $\dot M$, and $\chi^2_{\nu}=1.20$, 1.29 (522 d.o.f.) in the SS at the miniumum and maximum $\dot M$, respectively. The accretion rate in the hot phase is always larger than that in the clumps. 

We note the presence of an absorption feature at $\sim$4.4 keV in the HS MECS data, but not in either of the LECS data. This energy corresponds to an
Xe absorption line and it is most likely of instrumental origin (SDC, private
communication). Adding an absorption line at that energy (which we keep for the following fit) improves the fit to $\chi^2_{\nu} =0.99$, 0.96 (465 d.o.f.) at the minimum and maximum $\dot M$, respectively.

The two-phase model fits the SS data slightly worse than the HS ones. The SS $\chi^2_{\nu}$ is relatively large mainly due to the residuals around 1.6 keV. By including a narrow line at that energy into the model, we were able to considerably improve the fit quality. The minimum and maximum $\dot M$ $\chi^2_{\nu}$ become 0.95 and 1.06, respectively (519 d.o.f.). We
suspect that that line could be produced by Fe {\sc xxiii}. However, since we have no certainty about its origin, we leave it out of our final model given below.

\subsection{Compton reflection, the Fe K line and the final models}
\label{reflection}

The HS model is extremely reflection dominated, with $\Omega/2\upi$ reaching 50, which can be achieved if the X-ray source is obscured and we only see the light
reflected from some optically-thick medium. (Note that in our models, the wind is illuminated by the spectrum fitted in Section \ref{s:phenomenon}, which is also reflection-dominated.) Compton reflection should also give rise to an Fe K$\alpha$ line, which is neither included in the {\tt eqpair} model nor its presence is apparent in the residuals. Its equivalent width with respect to the reflection spectrum is $\sim$1 keV (\.Zycki \& Czerny 1994) for hard incident spectra, with the photon index $<2$. Therefore, we add a Gaussian line to the model and fit again the data but imposing the condition that the equivalent width, EW = 1 keV, with respect to the reflection component. This lowers the best-fit value of $\Omega/2\upi$. This model has the form,
\begin{eqnarray}
\lefteqn{F(E) = }\\
\lefteqn{\quad A_0(E) \left\{ A_{\rm w,c}(E) A_{\rm w,h}(E) \left[ F_{\rm C}(E) + F_{\rm Fe}(E) \right] +  F_{\rm w,c}(E) + F_{\rm w,h}(E)\right\}.  \nonumber}
\label{model4}
\end{eqnarray}
Adding the line improves the fit significantly for the HS data, the F-test probability for the improvement being by chance (Bevington \& Robinson 1992) is $<3\times 10^{-6}$. (If the Fe line is allowed to have a free normalization, it changes only slightly with respect to the fixed EW = 1 keV.) Note that the best-fit line energy is $\sim$6.7 keV, and it is moderately broad, $\sigma_{\rm Fe}\sim 0.3$ keV. The fit parameters are given Table \ref{fitparam}, and Fig.\ \ref{bestHS} shows unfolded spectrum with the residuals. Fig.\ \ref{components}(a) shows decomposition into the model components (following the method described in Hannikainen et al.\ 2005). 

The same procedure in case of SS was not successful. It was not possible to
introduce an additional Fe line to the model (although its predicted equivalent width is $\sim$3 times lower because of the softness of the spectrum above 7 keV). Thus, we keep the model as described in Section \ref{two_phase} and plotted in Fig.\ \ref{fits2}(f). The best-fit parameters are given in Table \ref{fitparam}. Fig.\ \ref{components}(b) shows decomposition into the model components. In all of the models, the ionization parameter, $\xi$, was found to be $\ll 1$, consistent with neutral reflection. 

\begin{figure}
\centerline{\includegraphics[width=74mm]{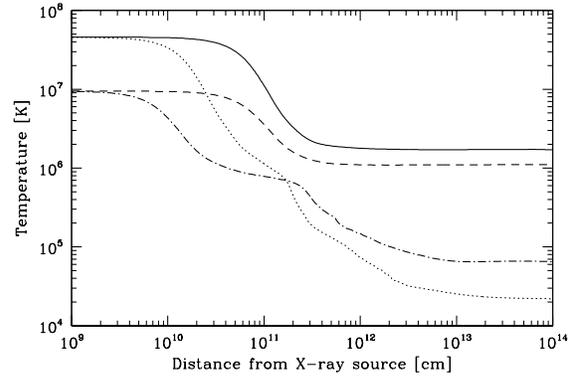}}
\caption{The wind temperature as a function of the distance from the X-ray source. The solid and dotted curves correspond to the homogeneous wind and the clumps, respectively, in the hard state. The dashed and dot-dashed curves correspond to the homogeneous wind and the clumps, respectively, in the soft state.
}
\label{wind_T}
\end{figure}

Fig.\ \ref{wind_T} shows the resulting wind temperature profiles as a function
of the distance from the X-ray source. We see the wind is strongly heated by
the X-ray source, achieving temperatures much higher than that of the stellar
radiation. Close to the X-ray source, the wind is at the Compton temperature,
equation (\ref{comptont}), i.e., very hot and almost completely ionized. Far
away, both the wind and photon density decrease proportionally to the square of
the distance. Thus, each phase reaches a constant ionization parameter,
$\xi_{\rm w}=L/(n r^2)$. Since the wind is then optically thin, its state is
determined only by $\xi_{\rm w}$ (Tarter, Tucker \& Salpeter 1969), and then
the temperature of each phase approaches constant, as seen in Fig.\
\ref{wind_T}. The wind Thomson optical depth along the line of sight is 0.31
and 0.43 in the HS and SS, respectively. Because {\tt cloudy} does not take
electron scattering into account in radiative transfer calculations, the resulting intrinsic spectra corrected for scattering will have higher fluxes (see Yaqoob 1997) than those given in Table \ref{fitparam}. While the effect is relatively small in our case ($\sim$10 per cent), it can be significant around the superior conjunction.  

\begin{table*}
\caption{The fit parameters of the best-fit models, equation (\ref{model3}), for the cases minimizing and maximizing $\dot M$. An additional Fe K$\alpha$ line from Compton reflection with the EW = 1 keV with respect to the reflection spectrum is included in the HS models. The electron temperature, $T_{\rm e}$, is calculated self-consinstently, and thus it is not a model parameter. The parameters marked with `f' were kept constant. $F_{\rm X}$ is the total flux in the unabsorbed model, and $F_{\rm X0}$ is the model flux without the reflection component. The uncertainties correspond to 90 per cent confidence for a parameter, i.e., $\Delta \chi^2=+2.71$. }
\begin{tabular}{lllll}
\hline
Parameter                          & HS$_{\rm min}$                         & HS$_{\rm max}$                          & SS$_{\rm min}$                     & SS$_{\rm max}$\\
\hline
$\dot M_{\rm h}\, [\msun$ yr$^{-1}$]      & $5.30^{+0.34}_{-0.22} \times 10^{-6}$  & $9.66^{+0.39}_{-0.35}\times 10^{-6}$    & $6.55^{+0.18}_{-0.18} \times 10^{-6}$ & $1.17^{+0.04}_{-0.03} \times 10^{-5}$ \\

$\dot M_{\rm c}\, [\msun$ yr$^{-1}$]      & $1.00^{+0.01}_{-0.01} \times 10^{-6}$  & $1.80^{+0.02}_{-0.01}\times 10^{-6}$    & $1.142^{0.002}_{0.002} \times 10^{-6}$      & $4.20^{+0.01}_{-0.01} \times 10^{-6}$ \\

$N_{\rm h}$                               & $0.31^{+0.04}_{-0.03}$                 & $0.38^{+0.04}_{-0.032}$                 & $0.47^{+0.03}_{-0.02}$                & $0.45^{+0.02}_{-0.03}$ \\

$N_{\rm c}$                               & $0.95^{+0.16}_{-0.17}$                 & $1.21^{+0.18}_{-0.12}$                  & $0.66^{+0.05}_{-0.05}$                & $0.60^{+0.03}_{-0.04}$ \\

$f$                                       & $0.01^f$                               & $0.01^f$                                & $0.005^f$                          & $0.005^f$ \\  

WR type                                   &  WNL                                   &  WNE                                    & WNL                                & WNE \\

velocity law                              &  (iii)                                   &  (i)                                     & (iii)                                & (i) \\

$M\, [\msun$]                             &  6.4                                   &  70                                     & 6.4                                & 70 \\

$\ell_{\rm h}/\ell_{\rm s}$               & $0.70^{+0.03}_{-0.02}$                 & $0.74^{+0.01}_{-0.01}$                  & $7.480^{+0.001}_{-0.085} \times 10^{-2}$ & $7.02^{+0.05}_{-0.07} \times 10^{-2}$ \\

$\ell_{\rm bb}$                           & $100^f$                                & $100^f$                                 & $100^f$                            & $100^f$ \\

$kT_{\rm bb}$ [keV]                       & $1.7^f$                                & $1.7^f$                                 & $1.2^f$                            & $1.2^f$ \\

$\ell_{\rm nth}/\ell_{\rm h}$             & $0.82^{+0.13}_{-0.04}$                 & $0.80^{+0.03}_{-0.03}$                  & $9.78^{+0.34}_{-0.08} \times 10^{-2}$ & $0.10^{+0.05}_{-0.04}$ \\

$\tau_{\rm p}$                            & $5.52^{+0.45}_{-0.28}$                 & $6.18^{+0.08}_{-0.06}$                  & $2.23^{+0.09}_{-0.01}$          & $1.81^{+0.12}_{-0.03}$ \\

$\Gamma_{\rm inj}$                        & $3.72^{+0.12}_{-0.07}$                 & $3.68^{+0.08}_{-0.11}$                  & $2^f$                              & $2^f$ \\

$\Omega/2\upi$                             & $10.0^{+0.82}_{-1.22}$                 & $11.5^{+0.77}_{-0.95}$                  & $10.93^{+0.26}_{-0.21}$            & $8.98^{+0.14}_{-0.14}$ \\

$kT_{\rm e}$ [keV]                        & $4.26$                                 & $3.96$                                  & $2.89$                             & $3.43$ \\

$N_{\rm C}$                             & $2.09^{+0.02}_{-0.04} \times 10^{-4}$  & $2.012^{+0.001}_{-0.009} \times 10^{-4}$  & $7.80^{+0.03}_{-0.03} \times 10^{-3}$ & $8.75^{+0.03}_{-0.04} \times 10^{-3}$ \\

$E_{\rm Fe}$  [keV]                       & $6.68^{+0.03}_{-0.03}$                 & $6.69^{+0.02}_{-0.03}$                  & --                                 & -- \\

${\sigma_{\rm Fe}}$ [keV]                 & $0.32^{+0.03}_{-0.03}$                 & $0.31^{+0.03}_{-0.03}$                  & --                                 & -- \\

$I_{\rm Fe}$ [cm$^{-2}$ s$^{-1}$]                                       & $1.52^{+0.08}_{-0.11}\times 10^{-2}$   & $1.51^{+0.07}_{-0.06} \times 10^{-2}$   & --                                 & -- \\

$F_{\rm X}$ [erg cm$^{-2}$ s$^{-1}$]    & $6.45 \times 10^{-9}$                  & $6.5 \times 10^{-9}$                    & $2.73 \times 10^{-8}$              & $2.96 \times 10^{-8}$ \\

$F_{\rm X,0}$ [erg cm$^{-2}$ s$^{-1}$]  & $2.66 \times 10^{-9}$                  & $2.71 \times 10^{-9}$                   & $2.50 \times 10^{-8}$              & $2.76 \times 10^{-8}$ \\
$\chi^2_{\nu}$                            & 0.92                                   & 0.91                                    & 1.18                               & 1.29\\
d.o.f.                                       & 463                                    & 463                                     & 522                                & 522 \\
\hline
\end{tabular}
\label{fitparam}
\end{table*}

\subsection{Cooling by the primary}
\label{cooling}

As the final step, we now estimate a possible change of the wind parameters caused by the WR stellar photons, see Section \ref{s:transfer}. Single WNE/WNL WR stars have $L_\star\simeq (1$--$30) \times 10^{38}$ erg s$^{-1}$ and the effective temperatures of $T_{\rm eff}\simeq (3$--$9) \times 10^4$ K (Hamann et al.\ 1995). To estimate the maximum cooling effect we add the blackbody of the brightest and coolest star ($L_\star =3\times 10^{39}$ erg s$^{-1}$, $T_{\rm eff}= 3\times 10^4$ K) to the intrinsic continuum assuming they both originate from the position of the compact object. This maximizes the effect of the blackbody radiation on the plasma in the line of sight. We then run the radiative transfer calculations to create emission and absorption models for the SS case with the maximum $\dot M$. 

We find that the inclusion of such strong cooling source influences only slightly the wind parameters. The $\dot M$ of the hot component decreases by only $\sim$11 per cent and becomes $9.3 \times 10^{-6}\msun$ yr$^{-1}$. The cool clumpy phase is not affected by the addition of blackbody photons and its $\dot M$ remains within its fit uncertainties given in Table \ref{fitparam}. Thus, not including the stellar blackbody in our calculations has a negligible effect on our results. 

\section{Discussion}
\label{discussion}

\subsection{The mass-loss rate}
\label{s:masslossrate}

An important result of this study is a new estimate of the mass loss rate independent of those of most of other Cyg X-3 studies. The $\dot M$ for either HS or SS has been found to be in the range of (0.6--$1.6)\times 10^{-5}\msun$ yr$^{-1}$. If the terminal wind velocity is different, $v_{\infty}'$, from that assumed by us, $v_{\infty}=(1.3$--$1.6)\times 10^8$ cm s$^{-1}$, our values of $\dot M$ should be rescaled by $\sim v_{\infty}'/v_{\infty}$. Our values agree with the usual range of values of $\dot M$ for WR stars of (0.5--$15) \times 10^{-5}\msun$ yr$^{-1}$ (Hamann et al.\ 1995). 
 
On the other hand, we disagree with Mitra (1996), who, similarly to us, estimated absorption in the WR star wind and claimed that for $\dot M > 1 \times 10^{-7}\msun$ yr$^{-1}$ the wind will be opaque to X-rays. He based that limit on some approximate considerations of the wind opacity, and we are unable to fully explain the discrepancy with our detailed quantitative results using the photoionization code {\tt cloudy}. A factor contributing (but not fully explaining) to this difference is his use of lower values for both $v_{\infty}$ and the separation between the stars. This leads to an increase of the density at the location of the compact object by an order of magnitude with respect to our calcuations. He did use the {\tt cloudy} model in a single case with a rather high $\dot M = 6.3\times 10^{-5}\msun$ yr$^{-1}$, for which his finding of the wind being completely opaque agrees with our results.

Kitamoto et al.\ (1995) attributed the orbital period change to the mass loss from the binary. They found, from this requirement, $\dot M=5.8 \times 10^{-7} (M/\msun)\msun$ yr$^{-1}$. For consistency with our results, $M\simeq (10$--$30) \msun$ is required. This includes a large range of possible masses of WR stars as well as allows for the presence of either a black hole or a neutron star.

Then, Waltman et al.\ (1996) found $\dot M \la 1.0 \times 10^{-5}\msun$ yr$^{-1}$ from the observed time delays of post-outburst jet peaks at lower frequencies with respect to higher frequencies. Such delays set limits
on the opacity of the stellar wind. Their limit overlaps with our allowed range of $\dot M$. However, their constraints can be partially relaxed if the wind is either very hot or cold and only partially ionized.

On the other hand, our estimates of $\dot M$ are significantly lower than those of van Kerkwijk (1993) and Ogley, Bell Burnell \& Fender (2001). Van Kerkwijk (1993) have obtained $\dot M \simeq 4\times 10^{-5} \msun$ yr$^{-1}$ based on IR observations of the wind emission, following the method of Wright \& Barlow (1975). Ogley et al.\ (2001) observed IR emission in quiescent and flaring states and found $\dot M \simeq (4$--$30) \times 10^{-5} \msun$ yr$^{-1}$. We note that similar discrepancies between the $\dot M$ calculated with different methods have been already previously observed in both single and binary massive systems (Moffat \& Robert 1994; Nugis, Crowther \& Willis 1998; Schild et al.\ 2004). The source of the discrepancy is the usual assumption of a smooth wind (Wright \& Barlow 1975) when calculating the wind IR excess emission. The excess is produced by two-body processes, in particular free-free and free-bound, which both the emission and absorption are enhanced in a clumpy wind by $1/f$ with respect to the smooth flow, see Section \ref{s:transfer}. Then, we have found (using the formalism of Wright \& Barlow 1975 generalized to an arbitrary velocity profile) that the net effect of clumping on measuring the mass loss rate using two-body processes and assuming a smooth flow is an overestimate by $f^{-1/2}$. This result holds for the measured luminosity coming from either the entire wind region (if it is optically thin at the observed frequencies) or its outer part of about unit optical depth, and it is a generalization of the corresponding result of Schild et al.\ (2004). In the case of a two-phase wind, our mass loss rate corresponds to the emission from a smooth flow with the rate of $[\dot M_{\rm h}^2/(1-f)+\dot M_{\rm c}^2/f]^{1/2}$, see Section \ref{s:clumpy} below. Given our values of $f\la 10^{-2}$, our range of the values of $\dot M$ is consistent with van Kerkwijk (1993) and overlap with the lower limit of Ogley et al.\ (2001). 

An additional constraint on $\dot M$ will come from the form of the orbital modulation in Cyg X-3 in the HS and SS (shown in H08). In this work, we have not attempted such calculations. Some preliminary modelling of phase-resolved spectra was done in Szostek \& Zdziarski (2005).

\subsection{Clumpy and inhomogeneous winds}
\label{s:clumpy}

As discussed above, strong evidence for clumpy winds in WR and early type stars comes the discrepancy between estimates of $\dot M$ using the IR excess and other methods, which is explained by the enhancement of the two-body emission by $f^{-1/2}$ in a clumpy medium. Additonal evidence in single WR stars comes from random variability in their flux and polarization (Robert, Moffat \& Seggewiss 1991; L\'epine et al.\ 2000 and references therein). Another piece of strong evidence for the inhomogeneities comes from the form of the line wings created by frequency redistribution of line photons by electron scattering, with clumping reducing the relative contribution of these wings (Hillier 1991). 

The inhomogeneities can be described as consisting of numerous dense clumps embedded in a rarefied interclump medium. Ions of low and moderate ionization stages are found only in clumps while the interclump medium is highly ionized. The cause for formation of clumps appears to be hydrodynamical instabilities in radiatively-driven winds, which lead to shocks, density enhancements and rarefactions (e.g., Owocki, Castor \& Rybicki 1988).

The evidence for a clumpy wind in Cyg X-3 given in this work stems from the form of the X-ray spectra from this object, and it is independent of the findings discussed above. A homogeneous wind is found to be strongly ionized and giving rise to an emission excess at $\la 2$ keV, not present in the data. The excess disappears when the wind is assumed to be clumpy. On the other hand, we still need a hot phase to account for the observed Fe edge at $\sim$9 keV. The resulting two-phase medium has the relatively low mass fraction in the clumpy phase of $\sim$1/5--1/3.

The values of $f$ found by us, $\la 10^{-2}$, are smaller than those of other authors. Hillier \& Miller (1999) used in their calculation $f \sim 0.1$ for WNC (carbon sequence WR) stars. The line profiles fitting including electron line wings typically yield $f \sim 0.04$--0.25 (Morris et al.\ 2000). Schild et al.\ (2004) found $f \simeq 0.06$ for WR binary $\gamma^2$ Velorum. Hamann \& Koesterke (1998) obtained $f\sim 0.06$--0.25. 

We note, however, that those authors assumed empty space between the clouds. In our case, the intercloud medium is filled with a hot smooth plasma which provides most of the wind mass. Then, the density contrast between the two media is $>f$, and it is $\sim 0.03$. Furthermore, we find that for a medium with an arbitrary form of inhomogeneities, the enhancement of the wind two-body emission is not by $f^{-1/2}$ but by $f'^{-1/2}$, where $f'$ is given by
\begin{equation}
f'={\langle n\rangle^2\over \langle n^2\rangle}.
\label{f2}
\end{equation}
In a one-phase medium composed of clouds only, $f'=f$. However, for a two-phase medium,
\begin{equation}
f'={(\dot M_{\rm h}+\dot M_{\rm c})^2\over \dot M_{\rm h}^2/(1-f)+\dot M_{\rm c}^2/f},
\label{fprime}
\end{equation}
which, for our results, yields $f'\simeq 0.07$--0.3, in full agreement with the estimates of the clumping factor given above. 

We also point out that the WR star in Cyg X-3, if synchronized with the binary motion, rotates at a speed comparable to the breakup velocity (as noted by Ogley et al.\ 2001). Namely, its rotation is 0.12--0.35 of the breakup value for the range of WR models considered in Section \ref{minmax}. This may lead to the wind being compressed towards the equatorial plane, similarly to the case of Be stars (Bjorkman \& Cassinelli 1993). This will lead to an anisotropy of the wind, not included in our treatment. Furthermore, the wind will be focused towards the compact object (Friend \& Castor 1982). These density enhancement will affect estimates of $\dot M$ from the IR emission, and may be partly responsible for the high values obtained by van Kerkwijk (1993) and Ogley et al.\ (2001). On the other hand, our method of determination of $\dot M$ is sensitive to the density along the line of sight, and a wind enhancement away from it will be not accounted for in the method.

\begin{figure*}
\centerline{\includegraphics[width=110mm]{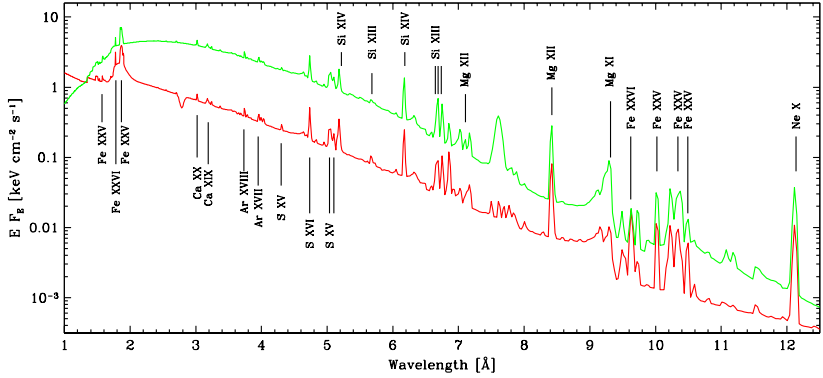}}
\caption{The best-fit models for the HS (lower, red) and the SS (upper, green) with line identification. The HS model includes the Gaussian line added in Section \ref{reflection}. The SS model also includes a line at $\simeq$7.6 \AA\ (1.6 keV), which we have been unable to identify. The absorption feature in the HS spectrum at $\simeq$2.8 \AA\ (4.4 keV) appears to be of instrumental origin.} 
\label{lines} 
\end{figure*}

\subsection{The line spectrum}
\label{s:lines}

In Fig.\ \ref{lines}, we plot the 1--12 \AA\ ($\sim$1--12 keV) SS and HS model
spectra and identify the strongest emission lines. We plot it in wavelength rather than in energy to facilitate a comparison with the results of {\it Chandra\/} high resolution spectroscopy (Paerels et al.\ 2000). 

We find that the lines resolved by {\it Chandra\/} are also the strongest lines of our model. However, recombination continua which were quite profound in the {\it Chandra\/} observation (in the HS) are not visible in the \sax\/ data.

\subsection{The model of the continuum}

At the best fits, our continua are dominated by Compton reflection. In the case of the HS, the the observed shape looks indeed remarkably similar to that of reflection. Recently, a HS Cyg X-3 spectrum measured by {\it INTEGRAL\/} was studied by H08, who also found a reflection-dominated as the best fit. We discuss below some physical implications of this model. 

The reflection component observed in the spectra of both XRBs and AGNs may originate in a cold torus surrounding the system or an accretion disc. In the case of XRBs, reflection from the companion is also possible, though its relative strength is usually small. The dominance of reflection means that we do not see directly the emitting source, which is obscured, but only its reflection from some medium. Such a situation is common in Seyfert 2 galaxies (e.g., Matt et al.\ 1996; Ikebe et al.\ 2000), where the central X-ray source is obscured by a large-scale torus, and we see the reflection from its back side. In XRBs, reflection may take place from a torus-like thickened inner accretion flow. A reflection-dominated spectrum has also been claimed from another XRB, GX 1+4 (Rea et al.\ 2005).

Then, Nandra \& George (1994) considered reflection from cold clouds embedded inside the hot plasma. However, the clouds have to be Thomson optically thick for this process to be effective. This is not the case in Cyg X-3, where we found that the total optical depth of the wind along the line of sight is $<1$ (Section \ref{s:cloudy}). The presence of clumps does not change the average optical depth (Osterbrock \& Flather 1959), and we can rule out this scenario. 

We have found that the reflecting material is close to neutral, and in particular, we see the absorption edge at $\sim$7 keV. On the other hand, we found that the Fe K line accompanying reflection is fitted at $\sim$6.7 keV, corresponding to highly ionized Fe. This is a serious problem for the reflection scenario. The observed 7-keV edge should be accompanied by a line at 6.4 keV. A line at that energy has been observed by {\it ASCA\/} (Kitamoto et al.\ 1994) and {\it Chandra\/} (Paerels et al.\ 2000). However, its equivalent width was very small, e.g., $\sim$60 eV in the latter (HS) data, much smaller than $\sim$1 keV predicted in the reflection-dominated scenario. If the observed 6.7 keV line represented the blue wing of the relativistic disc line with the rest energy at 6.4 keV, we should have also seen the K edge similarly blueshifted, but we do not.

We note that a remarkably similar puzzling situation is present in some Narrow-Line Seyfert-1 AGNs, e.g., 1H 0707--495 and IRAS 13224--3809 (Boller et al.\ 2002, 2003; Fabian et al.\ 2002, 2004; Tanaka et al.\ 2004; Gallo et al.\ 2004). Two classes of physical interpretations have been proposed, namely reflection and partial covering, both by a neutral medium. We consider them here in the context of Cyg X-3.

A reflector in the form of either cold torus or cold wind clumps in a natural
way produces a 7 keV absorption edge. It also should produce neutral iron
K$\alpha$ line but such line is excluded by the data. It has been argued that if the reflecting material is in the inner regions of the X-ray source and the reflection spectrum inluding the line is blurred by Doppler and gravitational redshift effects. The apparent edge at $\sim$7 keV is then the blue edge of the relativistically blurred iron emission line (Fabian et al.\ 2004). 

In the other Narrow-Line Seyfert-1 scenario, the source is partially covered by a thick cold absorber (Gallo et al.\ 2004). The partial absorber does not reduce the soft photon flux as much as a complete absorber would do, but, if the absorber is thick enough, it can leave its signatures at high energies and produce strong iron edge. In addition, similarly as in reflector scenario, if the absorber is located close to the compact object, the line will be blurred and thus invisible. Such (rather fine-tuned) scenarios appear in principle possible for Cyg X-3, but their detailed tests would require further studies.

On the other hand, the wind absorption is very strong in Cyg X-3 and the form of the spectra corrected for it is rather different from that observed, see, e.g., Fig.\ \ref{fits}. Indeed, we find it worrisome that the dominance of reflection is found by us in both the HS and the SS, whereas the accretion-flow geometry is, in general, dependent on the accretion rate, with an optically-thick disc thickening predicted at high rates (e.g., Shakura \& Sunyaev 1973; Jaroszy\'nski, Abramowicz \& Paczy\'nski 1980). We consider it quite possible that adding more complexity to our models would change our conclusion. In particular, a major driver of the enhanced reflection in the fit is the presence of the $\sim$ 7 keV edge in the data. That edge may be due to partial covering by some weakly ionized additional material, e.g., a structure at the outer edge of the accretion disc. 

We point out that the electron temperatures in the Comptonizing plasma in the HS obtained by us (Table \ref{fitparam}) are much lower than values typical to the hard state in either black-hole or neutron-star binaries. Also, the rather high nonthermal electron fractions found by us (Table \ref{fitparam}) are also not typical to the HS. These effects are likely to be due to the interaction of the very strong wind with the accretion disc, leading e.g.\ to shocks accelerating electrons out of the Maxwellian distribution, see a discussion in H08.

Thus, the constrains on the shape of the intrinsic continuum we have obtained in this paper are relatively limited. We consider our present combined wind and continuum study as a first step towards physical understanding of Cyg X-3.

\subsection{The luminosity}
\label{lum}

If we take the results of our spectral fits as a phenomenological approximation to the true intrinsic spectra, we can calculate the luminosities from the absorption-corrected X-ray bolometric fluxes, $F_{\rm X}$ (Table \ref{fitparam}), assuming isotropy. We also need to correct for the effect of the orbital modulation. Our fitted spectra are phase-averaged, and the peak fluxes are higher than the average ones by $\sim\! 1/0.75$ (H08). This yields $L\simeq (0.8$--$4)\times 10^{38}$ erg s$^{-1}$ (from the HS to SS) at $d=9$ kpc. This can be compared to the Eddington luminosity for a helium system of $2.5\times 10^{38}(M_{\rm X}/\msun)$ erg s$^{-1}$. For $M_{\rm X}= 20\msun$, the Eddington ratios are $\sim$2 and $\sim$8 per cent for the HS and SS, respectively, which are close to those seen in Cyg X-1 (e.g., Zdziarski et al.\ 2002), another high-mass binary wind accretor. This supports the black-hole nature of the compact object. We also point out that the two other known WR X-ray binaries, IC 10 X-1 and NGC 300 X-1, which have very similar X-ray luminosities of $\sim\! 10^{38}$ erg s$^{-1}$, both appear to contain black holes (Prestwich et al.\ 2007; Carpano et al.\ 2007a, b).

If the compact object is a neutron star, these Eddington ratios become $\sim$25--110 per cent, which are somewhat above those seen in the HS and SS of atolls, i.e., weakly-magnetized accreting neutron-star binaries (Gladstone, Done \& Gierli\'nski 2007). However, there are currently no known weakly magnetized neutron stars in high-mass X-ray binaries. The only hypothetical exception is the peculiar system Cir X-1; however, the supergiant nature of this companion suggested by Jonker, Nelemans \& Bassa (2007) appears relatively uncertain. On the other hand, the spectra of strongly-magnetized neutron-star binaries have an approximate bremsstrahlung (with $kT\sim 20$ keV) form and no detected X-ray disc blackbody emission (e.g., Apparao 1994), and thus they look completely different from the spectra of Cyg X-3.

On the other hand, if we take the intrinsic incident spectra implied by our reflection best-fit models as corresponding to physical reality, the assumption of the isotropy has to be dropped as we are now looking at the source at a particular inclination at which most of the intrinsic flux is not seen. Then, $L_{\rm X}\ga 4\upi d^2 F_{\rm X0} (\Omega/4 \upi)$, where $F_{\rm X0}$ is the observed unabsorbed flux of the direct component (Table \ref{fitparam}). The factor of $\Omega/4 \upi$ corresponds to the extreme situation in which the actual reflector solid angle is $\sim 4\upi$, which assumption minimizes the luminosity. This yields $L_{\rm X}\ga (1.7$--$16)\times 10^{38}$ erg s$^{-1}$. This corresponds to a relatively high Eddington ratio in the HS, implying, e.g., a high black-hole mass, see a discussion in H08. (Note that our incident spectrum in the HS is much harder at $E\la 20$ keV than that fitted by H08 for their reflection-dominated model, which results in a higher model luminosity than that found by us.) The presence of a neutron-star is unlikely as the Eddington ratio in the SS would be then $\ga$5.

\section{Conclusions}
\label{conclusions}

There are a number of essential features of our model which are new in
the history of the studies of Cyg X-3. We have used a realistic model of the WR
star, with a heavy small inner core and dense wind, and with the abundances typical to helium stars. We have used an accurate model of Comptonization together with accurate radiative transfer calculations to model the intrinsic X-ray emission in the vicinity of the compact object reprocessed by the surrounding wind. We have used high quality \sax\/ data, which are characterized by both relatively good spectral resolution and broad-band coverage.

We have used the X-rays emission as a probe of the wind properties. We have determined the wind mass-loss rate to be (0.6--$1.6)\times 10^{-5}\msun$ yr$^{-1}$, which is consistent with mass-loss rates of WR stars. We have disproved a previous claim that if Cyg X-3 contains a WR star its wind would be completely opaque to X-rays. We have also determined the structure of the wind illuminated by the X-rays, and found it contains both a hot phase (containing most of its mass) and cold dense clumps, occupying less than 1 per cent of the wind volume. The hot phase was needed to account for the strong $\sim$9 keV edge seen in the data, and the cold phase was needed to absorb a significant soft-excess emission produced by the hot phase. We have also found that the density enhancement in the clumps averaged over both phases is consistent with the clumping factor found in WR stellar winds using other methods. The lines emitted by the wind found in the \sax\/ data are consistent with the high-resolution measurement by {\it Chandra}. 

Our fits imply that the observed X-ray continua contain very strong Compton reflection components. This would point out to the presence of a geometrically thick torus in the accretion flow, which would obscure most of the intrinsic X-ray source but allow us to see its reflection. The reflecting material is found to be weakly ionized (from the presence of an edge at $\sim$7 keV), but no associated fluorescent Fe K$\alpha$ (at 6.4 keV) is found. This presents us with a puzzle remarkably similar to that encounted in a number of Narrow-Line Seyfert 1 galaxies. On the other hand, the finding of the dominance of reflection may be an artifact of not including enough details in our spectral modelling. In particular, an additional partial covering by a weakly ionized medium could be responsible for the $\sim$7-keV edge in the data. Based on the bolometric luminosity of the inferred intrinsic continua, the compact object in Cyg X-3 most likely a black hole. 

\section*{Acknowledgments}
We thank C. Done, M. Gierli\'nski, P. Lubi\'nski, and, especially, L. Hjalmarsdotter, for valuable discussions. This research has been supported in part by the Polish MNiSW grants 1P03D01128 and NN203065933 (2007--2010), and the Polish Astroparticle Network 621/E-78/SN-0068/2007.

\bsp

\label{lastpage}

\end{document}